\documentclass[12pt,preprint]{aastex}

\slugcomment{Kim-Demarque-Yi}

\shortauthors{Kim et al.}
\shorttitle{The $\alpha$-Enhanced $Y^2$ Isochrones}

\begin{document}

\title{The Y$^2$ Isochrones for $\alpha$-element Enhanced Mixtures}

\author{Yong~-Cheol Kim}
\affil{Department of Astronomy, Yonsei University, Seoul 120-749, Korea \\
kim@galaxy.yonsei.ac.kr}

\author{Pierre Demarque}
\affil{Yale University, Department of Astronomy, PO Box 208101, New Haven,
CT 06520-8101, USA \\ demarque@astro.yale.edu}

\author{Sukyoung K. Yi}
\affil{University of Oxford, Astrophysics, Keble Road, Oxford OX1 3RH, UK
\\ yi@astro.ox.ac.uk}

\and
\author{David R. Alexander}
\affil{Physics Department, Wichita State University, Wichita, KS 67260, USA
\\ dra@twsuvm.uc.twsu.edu}

\begin{abstract}
 
	We present a new set of isochrones in which the effect of
the $\alpha$-element enhancement is fully incorporated. 
	These isochrones are an extension of the already published set of 
Y$^2$ Isochrones (Yi et al. 2001: Paper~1), constructed for the 
scaled-solar mixture.  
	As in Paper~1, 
helium diffusion and convective core overshoot have been taken into account.
	The range of chemical compositions covered is 
$ 0.00001 \le Z \le 0.08 $.
	The models were evolved from the pre-main-sequence stellar birthline 
to the onset of helium burning in the core.
	The age range of the full isochrone set is 0.1 -- 20\,Gyr, while
younger isochrones of age 1 -- 80\,Myr are also presented up to the 
main-sequence turn-off.
	Combining this set with that of Paper~1 for scaled-solar mixture 
isochrones, we provide a consistent set of isochrones which can be used 
to investigate populations of any value of $\alpha$-enhancement.
We confirm the earlier results of Paper~1
that inclusion of $\alpha$-enhancement effects further
reduces the age estimates of globular clusters by 
approximately 8\% if [$\alpha$/Fe]=+0.3. It is important to 
note the metallicity dependence of the change
in age estimates (larger age reductions in lower metallicities). 
This reduces the age gap between the oldest metal-rich and metal-poor 
Galactic stellar populations and between the halo and the 
disk populations.
	We also investigate whether the effects of $\alpha$-enhancement 
can be mimicked by increasing the total metal abundance in the manner 
proposed by Salaris and collaborators. 
	We find such simple scaling formulae are valid at low metallicities
but not at all at high metallicities near and above solar.
	Thus it is essential to use the isochrones rigorously computed 
for $\alpha$-enhancement when modeling metal-rich populations, such as
bright galaxies.
	The isochrone tables, together with interpolation routines 
have been made available via internet.

\end{abstract}

\keywords{globular clusters:general -- stars:abundances -- stars:evolution -- stars:interiors}

\section{Introduction}

	This paper presents an extension of the $Y^2$ Isochrones: scaled-solar 
mixtures (Yi et al. 2001: Paper~1).
	Its purpose is to study the effect of the $\alpha$-element 
enhancement on theoretical isochrones.  
	Already for some time, it has become clear that the abundances of
some chemical elements, in particular the elements synthesized by nuclear 
$\alpha$ capture reactions (the so-called $\alpha$-elements; 
e.g., O, Ne, Mg, Si, S, Ca, and Ti etc.), are enhanced with respect to iron 
in Population II stars.
	\citet{Nor01} is one of the latest such studies which report that 
metal-poor stars do not have a scaled-solar chemical composition.
	One popular interpretation for this is that Population II stars
formed in a relatively fast collapse of the proto-galaxy, in which
a rapid chemical 
enrichment and $\alpha$-enhancement through efficient
feedback from massive stars took place.
	Similar $\alpha$-element enhancements have also been observed 
in super metal-rich (greater than solar) stellar populations. 
	Spectrophotometric studies of giant elliptical galaxies
\citep{Pel90, Rich92}, and more recent spectroscopic studies of stars in the 
bulge \citep{McW99} and thick disk \citep{Pro00} of the Galaxy, using the 
Keck~I telescope, indicate that various degrees of $\alpha$-element 
enhancement are commonplace.  
	A thorough discussion of the significance of stellar abundances 
in understanding the evolution of stellar populations 
can be found in the reviews by \citet{Whe89} and \citet{McW97}.
	Applications to elliptical galaxies, which also appear to be
$\alpha$-enhanced, have been performed by 
\citet{Wei95} and \citet{SW98}, 
using the spectral indices of \citet{Wor92}.

	Because opacity tables for
$\alpha$-enhanced mixtures covering the complete relevant range of
temperature and density became available only recently, 
several sets of theoretical isochrones have been 
constructed assuming scaled-solar abundances for the elements heavier 
than helium.
	However, the obvious limitation is that they refer to the 
scaled-solar compositions only.
	For stars with $\alpha$-enhanced chemical compositions, one needs 
then to construct stellar models and isochrones for the particular
distribution of metals.  
	Although it is possible in principle to calculate a large number of
possible combinations of $\alpha$-enhanced ratios, it would not be
practical considering the degree of the uncertainties in abundance analysis
and the size of the computational effort.
	\citet{Van00}, and \citet{Salas00} are among the latest who published
sets of the $\alpha$-enhanced isochrones assuming a particular choice for
the $\alpha$-enhanced chemical compositions, based on the OPAL opacities 
(Rogers \& Iglesias 1995, Iglesias \& Rogers 1996). 

	In this study, we construct a set of $\alpha$-enhanced isochrones 
which is fully consistent with the standard set of $Y^2$ Isochrones 
previously released for scaled-solar mixtures (Paper~1).
	The purpose of this research is twofold.  
	One is to provide an extensive and consistent set of the $Y^2$ 
Isochrones for general use.
	And, the other is to explore the validity of the common 
practice of mimicking $\alpha$-enhanced isochrones
by adopting more metal-rich non-$\alpha$-enhanced isochrones,
utilizing  a simple scaling formula in the heavy element abundance
\citep{Chi91, Cha92, Sal93}.

\section{Stellar Evolutionary Computations}

\subsection{The solar mixture}

	The original Anders-Grevesse study \citep{AG89} suggests 
a metal-to-hydrogen ratio $Z/X$ of 0.0267, based on meteoritic iron.
	The solar mixture by \citet{GN93} contains some significant
changes to the CNO abundance in comparison to the Anders-Grevesse value. 
	There was a major effort in the early 1990's to determine
new and more accurate CNO and Fe abundance in the Sun which were reported by
\citet{GN93}.  
	Finally, the solar iron abundance agrees with 
the meteoritic abundance; and in 1996, \citet{Gre96} quoted 
$(Z/X)_\odot = 0.0244 \pm 0.001$ at the surface of the Sun.
	However, more recent estimates suggest that systematic errors 
much larger than the errors quoted above may still exist in the measurements.
	\citet{GS98} lately found $(Z/X)_\odot = 0.0230$, with a 10\% error 
estimate, and \citet{Asp00}, using a solar model atmosphere constructed 
with a more realistic treatment of convection, derives $(Z/X)_\odot = 0.0226$. 
	This uncertainty in the observed $(Z/X)_\odot$ 
directly translates into the uncertainty in the theory-predicted efficiency 
of diffusion just below the convection zone, in the solar tachocline.  
	In fact, helioseismic inversions \citep{BA97} 
show that the tachocline region is the least well understood part of 
the standard solar model.

\subsection{The $\alpha$-element enhanced mixture}

 	Although the $\alpha$-elements are undoubtedly enhanced in some
stars, the enhancement factor for each $\alpha$-element is not well 
determined.  
	\citet{Van00} assumed a constant enhancement for most 
$\alpha$-elements in their work, while \citet{Salas00} took an 
empirically-guided enhancement.
	Considering the degree of uncertainty in the observation, however,
one cannot make a simple choice between these two assumptions.
 
	In this study, we have made the assumption of a constant enhancement,
as chosen by \citet{Van00}. 
	The scaled-solar abundance ration of metals are taken from 
\citet{GN93}.  
	In the $\alpha$-enhanced case, we adopt a mixture in which 
the $\alpha$-elements were enhanced by a constant factor with respect to 
the solar abundance ratios.
	The twice enhanced ([$\alpha$/Fe]=+0.3)and four times enhanced 
([$\alpha$/Fe]=+0.6) mixtures as well as the solar mixture
are summarized in Table \ref{tbl1}.  
	Note that, as in \citet{Van00}, we have assumed that 
the abundance changes in Mn and Al are anticorrelated with 
those of the $\alpha$-elements\citep{Van00,Nor01}.

\begin{deluxetable}{lccc}
\tablewidth{0pt}
\tablecaption{The assumptions for $\alpha$-enhancement
 \label{tbl1}}
\tablehead{\colhead{Element} &\colhead{[$\alpha$/Fe]=0.0}\tablenotemark{a} &
\colhead{[$\alpha$/Fe]=+0.3} &\colhead{[$\alpha$/Fe]=+0.6} }
\startdata
C & 8.55 &  & \\
N & 7.97 &  & \\
O & 8.87 & 9.17 &9.47 \\
Ne & 8.08 & 8.38 &8.68 \\
Na & 6.33 & 6.63 &6.93 \\
Mg & 7.58 & 7.88 &8.18 \\
Al & 6.47 & 6.17 &5.87 \\
Si & 7.55 & 7.85 &8.15 \\
P & 5.45 & 5.75 &6.05 \\
S & 7.21 & 7.51 &7.81 \\
Cl & 5.50 & 5.80 &6.10 \\
Ar & 6.52 & 6.82 &7.12 \\
K & 5.12 &  & \\
Ca & 6.36 & 6.66 &6.96 \\
Ti & 5.02 & 5.32 &5.62 \\
Cr & 5.67 &  & \\
Mn & 5.39 & 5.24 &5.09 \\
Fe & 7.50 &  & \\
Ni & 6.25 &  & \\
\enddata
\tablenotetext{a}{ The scaled-solar abundance ratios of metals are taken
 from \citet{GN93}.}
\tablecomments{The abundance of the elements in logarithmic scale,
$\log N_{el}/N_H +12$, where $N_{el}$ is the abundance by number. }
\end{deluxetable} 

\begin{figure}
\plotone{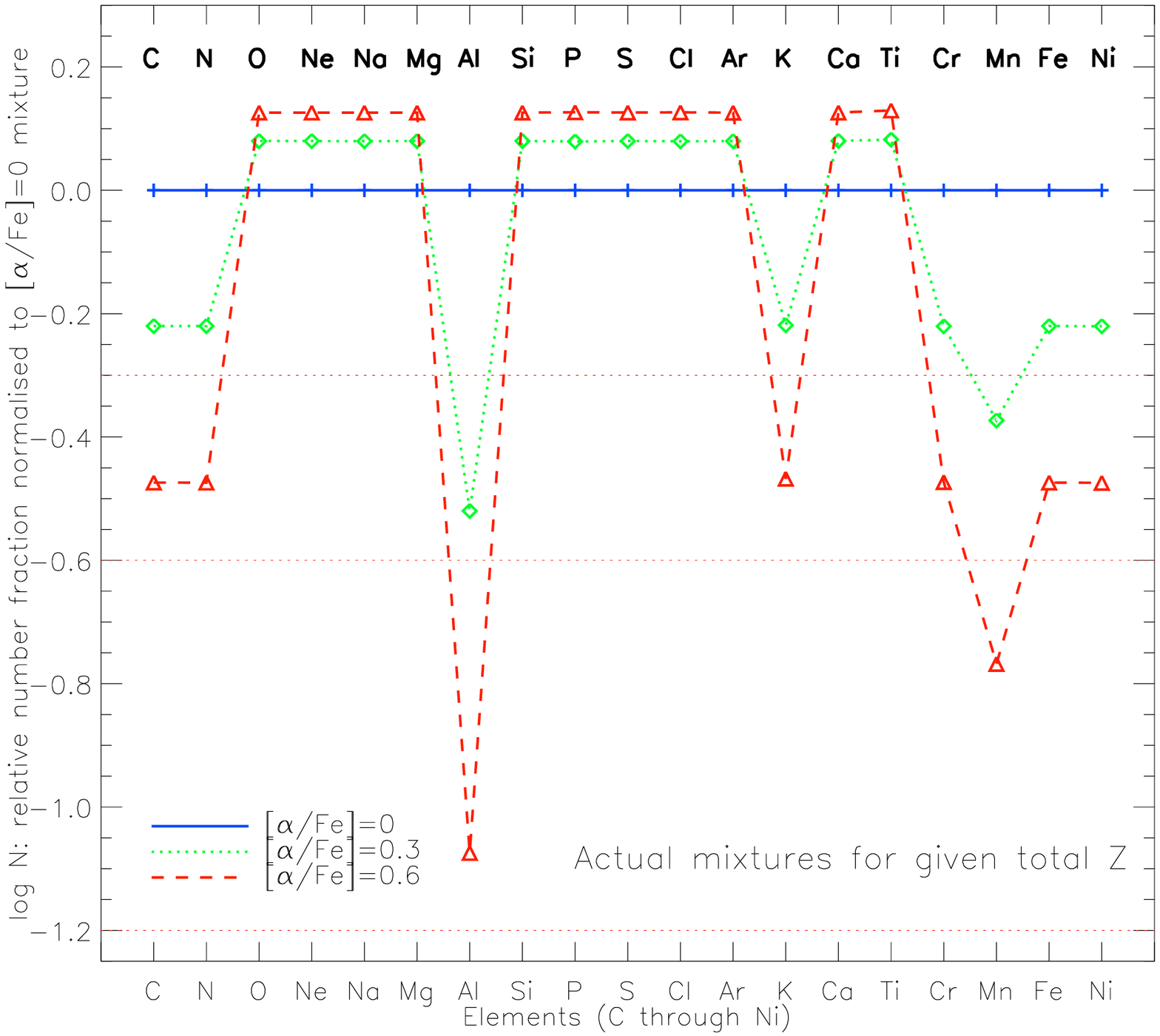}
\caption[f1.eps]{The relative number fractions of various
elements for fixed $Z$ for three values of $\alpha$-enhancement. 
Because the total metallicity $Z$ is fixed, the $\alpha$-enhanced mixture
is less abundant in non-$\alpha$-elements, such as Fe.
\label{f1}}
\end{figure}

Figure \ref{f1} shows the relative abundances in the two 
$\alpha$-element enhanced mixtures (twice solar and four times solar,
respectively), normalized to the solar mixture abundances.
The legends [$\alpha$/Fe]=0.0, 0.3, and 0.6 denote solar mixture without
$\alpha$-enhancement, 
two-times, and four-times $\alpha$-enhanced mixtures, respectively.
Note that because total $Z$ is fixed, $\alpha$-enhanced mixtures
are only slightly (approximately by 0.1) above the non-$\alpha$-enhanced 
mixture (top horizontal line at logN=0).
When all mixtures are normalized to the $\alpha$-elements (e.g., O, Mg)
the amount of $\alpha$-enhancement appears to be roughly a factor of two 
in the [$\alpha$/Fe]=0.3 mixture (dotted line) and four in 
the [$\alpha$/Fe]=0.6 mixture (dashed line), respectively.
As mentioned above, Al and Mn have been assumed to be depleted. 
It is important to remember that our isochrones have been 
computed for a given total metallicity $Z$.
	In this context, more $\alpha$-enhanced isochrones are simply 
less abundant in non-$\alpha$-elements.
Oxygen is the most important element, and it affects both energy generation 
and opacity in an important way.  Neon is also a significant contributor to
interior opacities near the main sequence.

\subsection{Metallicity parameters and composition scaling}
 
	Departing from the solar mixture can create confusion in assigning 
chemical abundance parameters to the isochrones, and in making comparisons 
with observations.  
	Depending on the kind of data available, it is 
at times useful to parameterize heavy element content in terms of [Fe/H], 
or at other times in terms of $Z$.  
	Generally, [Fe/H] is the quantity measured observationally by 
spectroscopy or photometry. 
	Relative abundances by number are also favored in theoretical stellar 
atmosphere models. 
	To help clarify possible confusion, Table \ref{tbl2} shows the 
conversion from [Fe/H] to $Z$ for the solar mixture and the 
two $\alpha$-enhanced mixtures used in the calculations.
	Table \ref{tbl2} was constructed assuming that 
$\Delta Y/ \Delta Z=2$ for the chemical enrichment.

\begin{deluxetable}{cccc}
\tablewidth{0pt}
\tablecaption{Conversion from [Fe/H] to $Z$ \label{tbl2}}
\tablehead{ \colhead{[Fe/H]}  &
\multicolumn{3}{c}{$Z$}\\ & \colhead{[$\alpha$/Fe]=0.0} &
 \colhead{[$\alpha$/Fe]=+0.3}  & \colhead{[$\alpha$/Fe]=+0.6} }
\startdata
-3.0 & 0.000019 & 0.000032 & 0.000058\\
-2.5 & 0.000062 & 0.000102 & 0.000182\\
-2.0 & 0.000195 & 0.000321 & 0.000574\\
-1.5 & 0.000615 & 0.001012 & 0.001807\\
-1.0 & 0.001935 & 0.003174 & 0.005627\\
-0.5 & 0.006021 & 0.009774 & 0.016990\\
 0.0 & 0.018120 & 0.028557 & 0.047000\\
 0.5 & 0.049711 & 0.072793 & 0.106471\\
 1.0 & 0.110798 & 0.142689 & 0.177489\\
\enddata
\end{deluxetable}

	There is another motivation for studying the relationship 
between $Z$ and [Fe/H] for different chemical mixtures.  
	Until recently, interior opacity tables were only
available for the solar mixture, and in studies of very metal-poor systems, 
such as globular clusters and dwarf spheroidal galaxies, there was interest 
in mimicking the effects of $\alpha$-enhancement by correcting the 
metallicity $Z$ used in the opacities.  
	It was first pointed out by \citet{Chi91} 
--- see also \citep{Cha92, Sal93, SC96} --- that one can simply write:   
\begin{equation}
Z = Z_0 (0.694 f_\alpha + 0.306 )
\end{equation}
where $f_{\alpha}$ is the $\alpha$-element enhancement factor 
and $Z_{0}$ is the heavy element abundance by mass for the solar mixture 
with the same [Fe/H].
	Note that the coefficients used here are taken from \citet{paper1},
and are slightly different from those of Salaris et al. (1993) because they 
apply to the Grevesse-Noels solar mixture \citep{GN93}.
	Comparing with Table~2 reveals that this formula, which yields the 
correct values of $Z$ at extremely low metallicities, becomes 
increasingly less exact at higher metallicities. 
	This is due to two main reasons: at high metallicities,
(1) the heavy elements contribute a larger fraction of the total 
mass in the mixture relative to hydrogen and helium, and (2) the 
helium mass fraction $Y$ increases steeply with increasing $Z$ (Table~2 
assumes $\Delta Y/\Delta Z$=2).

	It is now possible to calculate opacities for $\alpha$-enhanced 
mixtures and investigate the validity of such scaling 
formulae for a wide range of metallicity.  
	Detailed comparisons between isochrones based on scaled-solar 
mixtures and isochrones constructed for the correct $\alpha$-enhancement 
are discussed in the rest of this paper.  
	We will confirm, as previously noted by several authors 
\citep{Cha92, Sal93, Van00}, that simple scaling formulae are adequate 
at low metallicities.  
	However, not surprisingly, such scalings and even simple 
interpolations turn out to be very unreliable at high metallicities.

\subsection{Input physics and opacities}

	In order to maintain as much homogeneity and 
consistency with our solar mixture isochrones as possible,
the input physics, calibrations and computational details were kept the 
same for this set of evolutionary tracks as in Paper~1.
	A full description can be found in \S~2 of Paper~1.

	The only difference is in the opacity tables.
The effect of the $\alpha$-element enhanced mixture
on the opacities has been taken into account.
	The OPAL opacities for the \citet{GN93} mixture were used for the 
computation, as in Paper~1.
	These opacities are now regarded as one of two best sets of 
Rosseland mean opacities available today, the Opacity Project tables being 
the other \citep{Sea97}. 
	The fact that both sets of opacities, which were calculated 
independently and using different physical models, are for the most part 
in excellent agreement, lends support to their validity.  
	Lately, the OPAL group released newer tables for the \citet{GN93} 
solar mixture that include the effects of seven additional heavy elements 
as well as some physics changes \citep{RI95,IR96}.  
	The OPAL interior opacities used in this paper were generated for 
each $\alpha$-element enhanced mixture from the OPAL web site 
({\tt www-phys.llnl.gov/Research/OPAL/}).

	At low temperatures, tables were computed specifically for the 
$\alpha$-enhanced mixtures by one of us (DRA).  
	These low temperature opacity tables were constructed using the 
same physics as the solar mixture low temperature opacity tables published by 
\citet{AF94}.

\subsection{Solar Calibration, Evolutionary Tracks and Isochrone Construction}

	For internal consistency, we used the same solar calibration as 
in Paper~1.  
	In Paper 1, a model of the Sun was evolved from the zero-age main 
sequence (ZAMS) to the current solar age of 4.55\,Gyr \citep{Gue89}. 
	The mixing length to scale height ratio ($l/H_{p}$) and the initial 
helium content ($Y_{ZAMS}$) were adjusted in the usual way so as to match the
solar luminosity, radius, and the observed solar ratio of heavy element to 
hydrogen abundance $Z/X=0.0244$ \citep{Gre96} at the solar age.
	The mixing length parameter $l/H_{p} = 1.7432$, the initial 
$X=0.7149$, and $Z=0.0181$ were used to produce this match.
	At the solar age, the solar model has the surface hydrogen abundance
$X=0.7463$, which satisfies $Z/X=0.0244$, consistently with the 
recent standard solar models of \citet{Gue97} and \citet{Win02}.
	The same mixing length parameter $l/H_{p}$ was used for all models. 

	Evolutionary tracks were constructed for the masses 0.4 to 
5.0$M_{\odot}$ and for the metal abundances, $ 0.00001 \le Z \le 0.1 $. 
	In this study, the $\Delta Y/ \Delta Z$  was assumed to be 2.0, 
which is consistent with the initial  $(Y, Z)_{0}$ = (0.23, 0.00) and the 
solar calibration above.
	Unlike many studies, which are based on evolutionary tracks with 
starting points on the ZAMS, all models in this grid were evolved from the 
pre-MS stellar birthline \citep{PS90,PS93} to the onset of helium burning.
	This enabled us to construct isochrones for younger ages.
	The age range of the isochrone set was  0.1 -- 20\,Gyr.  
	All isochrones in this age range are suitable for population synthesis.
	
	In addition, very young isochrones, for 
ages of 1, 2, 4, 6, 8, 10, 20, 40, 60, and 80\,Myr have been included.  
	However, because these isochrones 
reach only up to 5\,$M_{\odot}$, they do not cover the red giant branch (RGB). 
	So, one must not use them by themselves for population synthesis 
studies. 
	But they would be useful for color-magnitude diagram (CMD) 
fitting of young clusters, in 
particular as they begin at the pre-MS birthline.
	The treatment of convective core overshoot is the same as in Paper 1 
\citep{paper1}.  
	The overshoot parameter was chosen to be 0.2\,$H_{p}$ for 
age$<$3\,Gyr.
	The two independent color transformation tables of 
\citet{Gre87} and of \citet{Lej98} are adopted as
in Paper~1, making two separate sets of isochrones available.

\section{Effects of Enhanced $\alpha$-element Abundances}

\subsection{Isochrone morphology}

\begin{figure}
\plotone{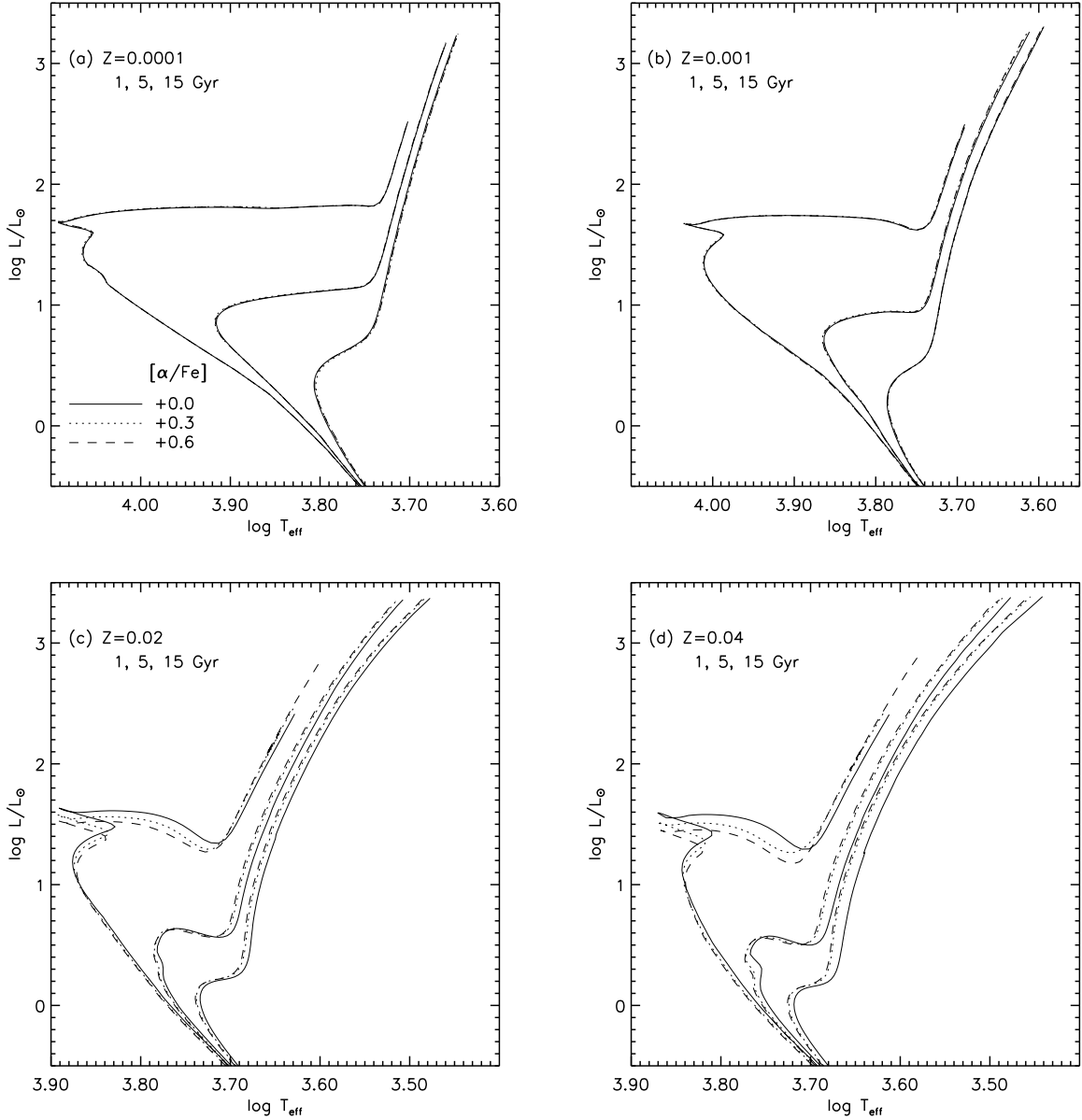}
\caption[f2.eps]{The effect of $\alpha$-enhancement on 1, 5, 15\,Gyr 
isochrones (from left to right) for four values of total metallicity $Z$. 
Continuous, dotted, and dashed lines are for [$\alpha$/Fe] = 0.0, 0.3, \& 
0.6, respectively. The effect of $\alpha$-enhancement is
simply an effect of metallicity increase at low metallicities.
At low metallicities, shown in panels (a) and (b), $\alpha$-enhanced isochrones
virtually overlap with scaled-solar models.
But the effect becomes very complex at metallicities greater than solar.
\label{f2}}
\end{figure}

\begin{figure}
\plotone{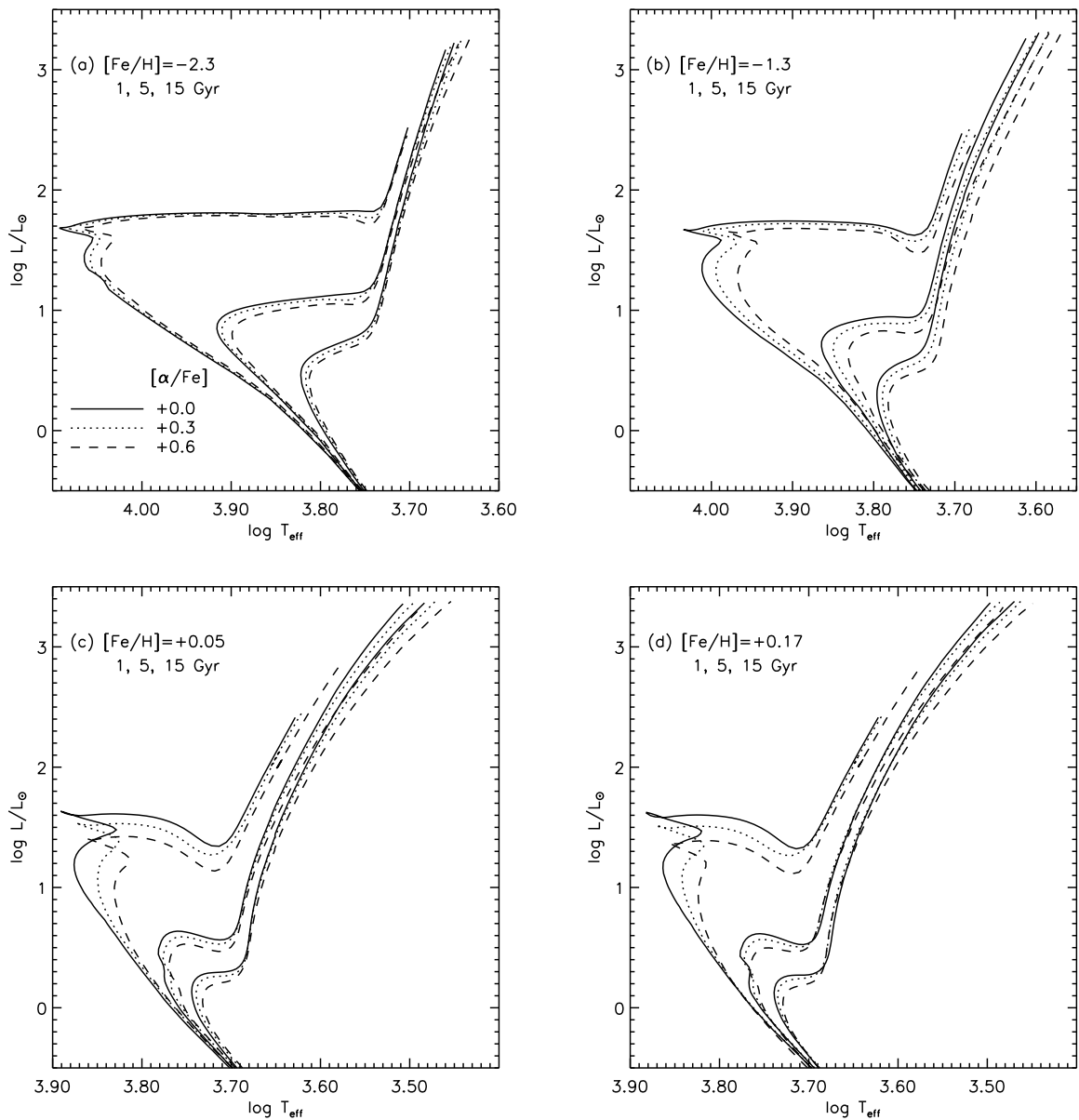}
\caption[f3.eps]{The effect of $\alpha$-enhancement on 1, 5, 15\,Gyr 
isochrones (from left to right) for four values of [Fe/H]. 
Continuous, dotted, and dashed lines are for [$\alpha$/Fe] = 0.0, 0.3, \& 
0.6, respectively. The effect of $\alpha$-enhancement is
simply an effect of metallicity increase at low metallicities.
But the effect becomes very complex at metallicities greater than solar.
\label{f3}}
\end{figure}

	The overall effects of the enhanced $\alpha$-element abundances 
on the isochrone morphology are complex at all metallicities because they 
involve a non-linear interaction between opacities and energy generation.  
	The effects are shown in Figures \ref{f2} and \ref{f3}.

Figure \ref{f2} shows that low metallicity isochrones (panels a and b)
 depend primarily
on the total amount of heavy elements, and not much on the mixture 
of heavy elements. 
	The effects of the $\alpha$-element enhancement are still complicated
since they affect both the energy generation and the opacity, but they are  
less noticeable in the HR-diagram. 
	The location of the giant branches and the turn-offs are not a 
monotonic function of the degree of $\alpha$-enhancement.
	Although these effects are very small, they are visible
in the figures.  
	For example, the MS turn-off temperatures of the young (1\,Gyr) 
isochrones are not in linear relation with the magnitude of 
$\alpha$-enhancement at extremely low metallicities (panel a). 
	These non-linear behaviors of the isochrone morphology are the first 
indications that a simple scaling law cannot fully reproduce the 
$\alpha$-enhancement effects.

	The complexity of $\alpha$-enhancement effects are magnified in 
metal-rich isochrones (Figure \ref{f2}-(d) for example).
	When the total metal abundance, $Z$, is fixed, increased
[$\alpha$/Fe] means `less' iron abundance which is the main provider of
free electrons. 
	Near the turn-off, this results in the $\alpha$-enhanced 
isochrones behaving as if they have lower metal abundance in comparison 
to the scaled-solar ones.
	As shown in this figure, the $\alpha$-enhanced isochrones generally 
have their turn-offs at higher temperature.  
	For the younger isochrones whose turn-off stars
derive most of their luminosity from the CNO cycle during the 
main sequence (MS) phase, $\alpha$-enhancement also affects the MS lifetime.
	Thus, the turn-off masses of the $\alpha$-enhanced isochrones are 
smaller, resulting in the lower luminosity and higher temperature
of the turn-offs as the $\alpha$-element abundance increases.
	While the location of the MS is mainly affected by the oxygen 
abundance, that of the red giant branch is mainly determined 
by the iron abundance and the abundance of $\alpha$-elements
heavier than oxygen. 
	For a given $Z$, more $\alpha$-elements means less iron,
which in turn means fewer free electron, and lower $H^{-}$ atmospheric opacity.
	Thus, as [$\alpha$/Fe] increases, the red giant branch of the 
isochrones shift toward higher temperature and luminosity.

	When the isochrones are plotted for a fixed [Fe/H], as in 
Figure \ref{f3}, the behavior of the isochrone morphology is easier 
to understand.  
	The major factor here is the overall metal abundance $Z$.  
	For a fixed [Fe/H], the $\alpha$-element enhancement means simply 
the increase of the overall metal abundance. 
	The clearest linear trend is visible in Figure \ref{f3}-(b)
where [Fe/H]=$-1.3$ (typical Galactic globular cluster metallicity) models
are displayed.
	Here again, however, one may find a clue that simple scaling laws 
cannot be valid: 
although one notices a general trend toward
lower temperature and luminosity with increasing [$\alpha$/Fe], 
changes in the gradient of the giant branches with increasing luminosity 
are recognizable even in the 
metal-poor cases. 

	At the lowest metallicities, the main source of opacity in the 
interior is derived from the free-free transitions of hydrogen and helium, 
and models on the MS are unaffected by the relative heavy element abundances.  
	The principal effect of $\alpha$-enhancement is to increase the 
relative importance of the CNO cycle in the energy generation. 
	As first pointed out by \citet{SI68}, this results in an 
earlier central hydrogen exhaustion, resulting in a less luminous turn-off 
for a given age.  
	This effect is important in that CNO enhancement with respect to 
hydrogen reduces the derived ages of globular clusters with a given 
[Fe/H] without modifying appreciably the MS position (Figure 6).
	At higher [Fe/H], the interplay between opacity effects and 
energy generation is more complex, as is evident in Figure \ref{f3}-(d).

\begin{figure}
\plotone{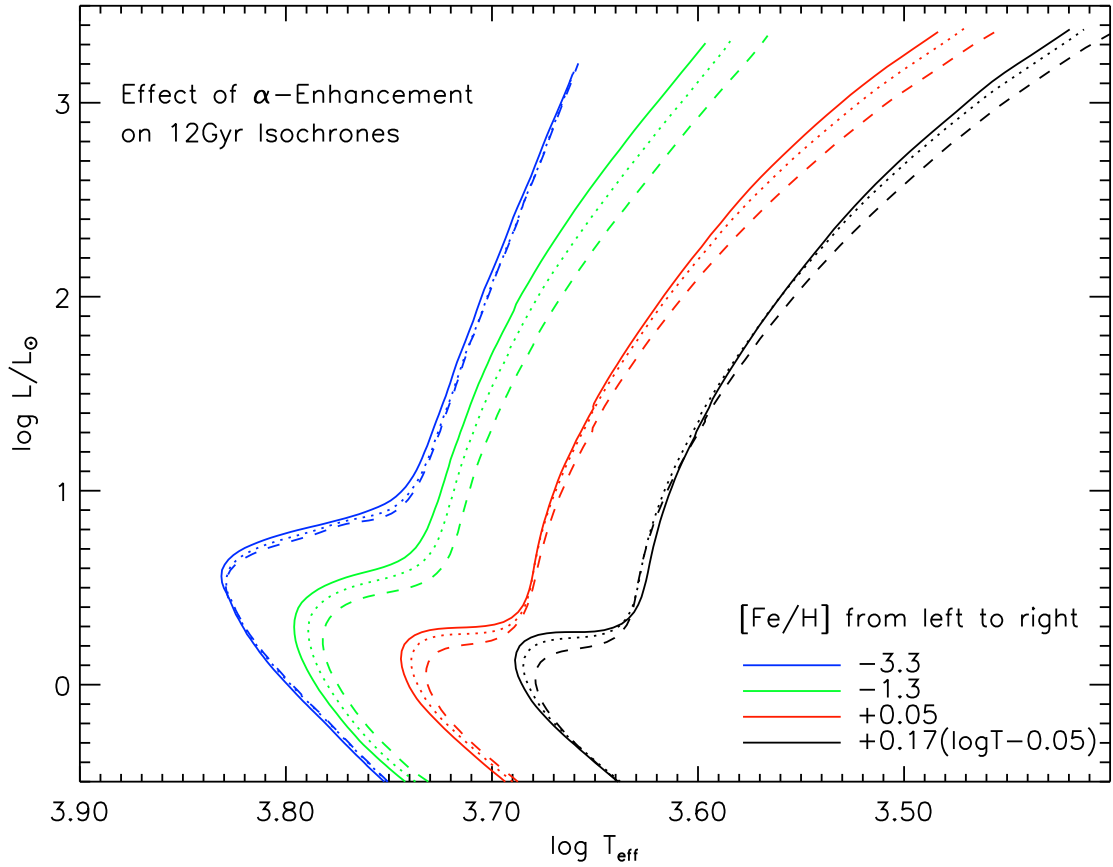}
\caption[f4.eps]{The effect of $\alpha$-enhancement on 12\,Gyr 
isochrones for a wide range of values of [Fe/H]. Continuous, dotted, and
dashed lines are for [$\alpha$/Fe] = 0.0, 0.3, \& 0.6, respectively.
The [Fe/H]=+0.17 isochrones are shifted to the right by 0.05 in $\log T_{\rm eff}$.
Note that the effect is highly complex and non-linear at high metallicities.
\label{f4}}
\end{figure}

	The complex metallicity dependence of the effects 
of $\alpha$-enhancement is illustrated 
in Figure \ref{f4}.
	The four sets of 12\,Gyr isochrones are for various values of
[Fe/H] and $\alpha$-enhancement.
	Note that the most metal-rich isochrones (a group of three on the
right) have been shifted by $\log T_{\rm eff} = 0.05$ to 
the right so that they appear 
separated from the [Fe/H]=0.05 (approximately solar) isochrones.

\subsection{Isochrone Ages}

	The changes in age estimates from the twice and the four times 
$\alpha$-enhanced cases with respect to that of the Revised Yale Isochrones
(Green et al. 1987), are summarized in Table \ref{age}, together with 
that of the updated input physics (the $Y^{2}$ Isochrones--solar mixture).
      The ages were derived based on the turn-off luminosity defined as
(1) total theoretical luminosity (denoted as `$L$'), (2) visual 
magnitude estimated using the Green et al. table, $V$(G), and (3) 
visual magnitude estimated using the Lejeune et al. color table, $V$(L).

\begin{figure}
\plotone{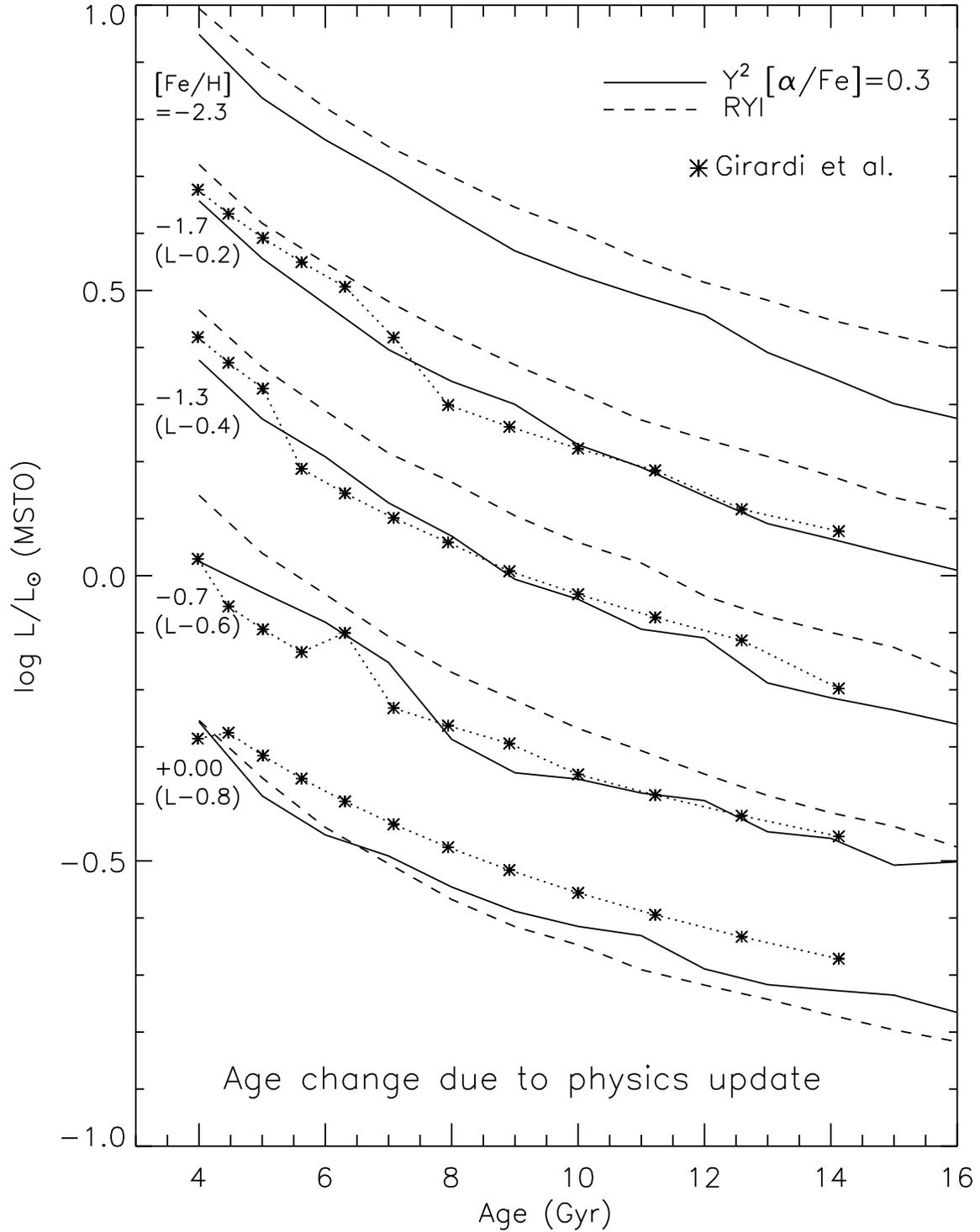}
\caption[f5.eps]{
The effect of physics update on the
estimation of the age, using isochrones of various values of [Fe/H]. 
The old Revised Yale Isochrones and Padova group's isochrones 
\citep{Gir00} are not $\alpha$-enhanced. For the purpose of comparison, 
we show the $Y^{2}$ [$\alpha$/Fe]=0.3 models as well.
As in Figure 14 of Paper 1, each metallicity group of three models
has been vertically displaced by a certain amount (number in parenthesis).
For typical Milky Way globular cluster metallicities, approximately 20 to 23\%
age reduction has been made by the physics update.
\label{f5}}
\end{figure}

	The age changes in the [$\alpha$/Fe]=0.0 `$L$' column are the
same as those in Table 7 of Paper~1.
	The approximately 15\% age reduction in Galactic globular cluster
metallicity range comes from the update in the physical input
in stellar model calculation and does not include the effect of 
$\alpha$-enhancement.
	Quite expectedly, the two  $\alpha$-enhanced models (columns 5 and 8) 
show that there is further age reduction when $\alpha$-enhancement is 
considered.
	But it should be noted that here we are not comparing the models
of the same total $Z$ any more.
	Instead, we have fixed [Fe/H], which is usually the observationally
measured quantity, and inspected the effects by changing [$\alpha$/Fe],
which is similar to increasing total $Z$.
	Note that the $\alpha$-element enhancement would not affect much 
age estimates if the total $Z$ were to be fixed.
	The amount of further age reduction due to $\alpha$-enhancement
is approximately 8 and 23\% for [$\alpha$/Fe]=0.3 and 0.6 cases, 
respectively.
	The total age reduction, derived from the turn-off luminosity,
between RYI and $Y^{2}$ Isochrones is as much as 19--23\% for moderately 
$\alpha$-enhanced populations. 
	Figure \ref{f5} shows this effect using
[$\alpha$/Fe]=0.3 isochrones.

\begin{figure}
\plotone{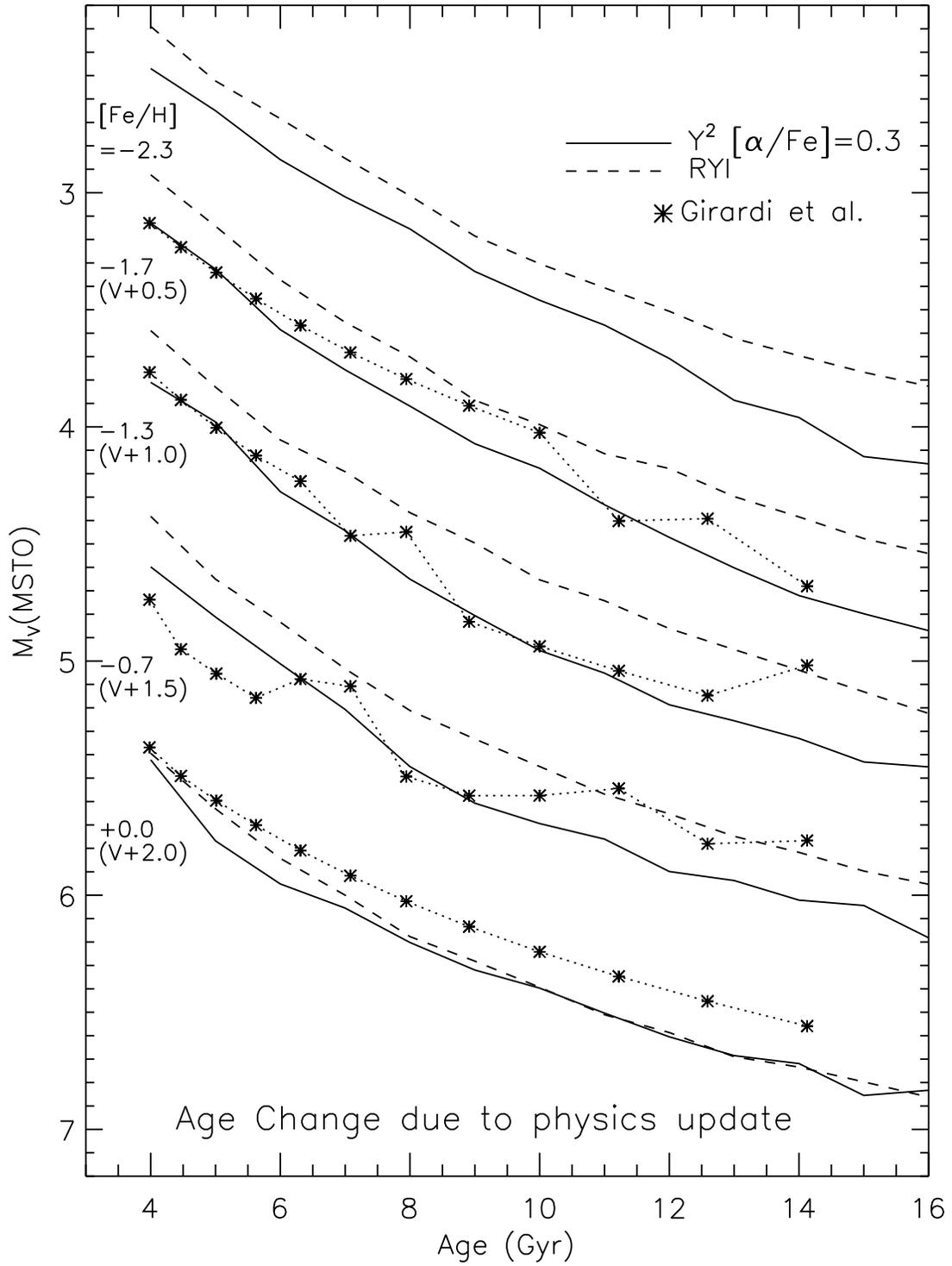}
\caption[f6.eps]{
Same as Figure \ref{f5}, but turn-off $V$ magnitudes 
(based on Lejeune et al. color table) have been used.
\label{f6}}
\end{figure}

	We prefer theoretical luminosities ($L$) for this analysis 
to magnitudes ($V$) because $L$ is free from the uncertain 
luminosity-to-magnitude conversion scheme, 
which is a part of the color transformation procedure.
	Yet, observers depend on observed quantities;
and thus we present similar comparisons using turn-off $V$ magnitudes
in other columns of Table \ref{age} and in Figure \ref{f6}.
	The $V_{TO}$-based age reduction is by and large 
comparable to that from theoretical luminosity.
	All this illustrates the difficulty of estimating 
stellar ages in systems for which only [Fe/H], but not [$\alpha$/Fe], 
is known.

\begin{deluxetable}{cccccccccc}
\tablewidth{0pt}
\tablecaption{Change in percent of turn-off-luminosity-based age estimate from RYI to $Y^{2}$ \label{age}}
\tablehead{
 &\multicolumn{3}{c}{[$\alpha$/Fe]=+0.0}&
\multicolumn{3}{c}{[$\alpha$/Fe]=+0.3}&
\multicolumn{3}{c}{[$\alpha$/Fe]=+0.6}\\
\colhead{[Fe/H]} & \colhead{$L$\tablenotemark{a}} & 
\colhead{$V$(G)\tablenotemark{b}} & \colhead{$V$(L)\tablenotemark{c}} &
\colhead{$L$\tablenotemark{a}}& \colhead{$V$(G)\tablenotemark{b}} &
\colhead{$V$(L)\tablenotemark{c}} & \colhead{$L$\tablenotemark{a}} &
\colhead{$V$(G)\tablenotemark{b}}&\colhead{$V$(L)\tablenotemark{c}}}
\startdata 
-2.3    &-15&-12&-12    &-19&-18&-18    &-31&-25&-26\\
-1.7    &-14&-12&-13    &-23&-21&-21    &-40&-32&-35\\
-1.3    &-15&-13&-16    &-23&-20&-26    &-38&-35&-40\\
-0.7    &-10&-6 &-14    &-19&-12&-20    &-26&-26&-35\\
 0.0    & 12& 14&  6    & 7&  6 &-3     &-8& -8 &-18\\
\enddata

\tablenotetext{a}{The ages were derived from the theoretical 
	turn-off luminosities.}
\tablenotetext{b}{The ages were derived from the V magnitude of the turn-off
	based on the Green color tables.}
\tablenotetext{c}{The ages were derived from the V magnitude of the turn-off
	based on the Lejeune color tables.}
\end{deluxetable}

\section{Scaling formulae that simulate $\alpha$-enhancement effects}

	Many studies have shown that
the effects of the $\alpha$-enhancement on the early evolution
of low mass stars may be treated by increasing the overall $Z$
of models computed for scaled-solar mixtures of the heavy elements
\citep{Chi91,Cha92,Sal93,SC96}.
	\citet{Sal93} type formula for the \citet{GN93} mixture has
been mentioned above as Eqn.~(1).
	The validity of such a scaling procedure should be investigated
especially for the higher metallicities \citep{Wei95,SW98}.

\begin{figure}
\plotone{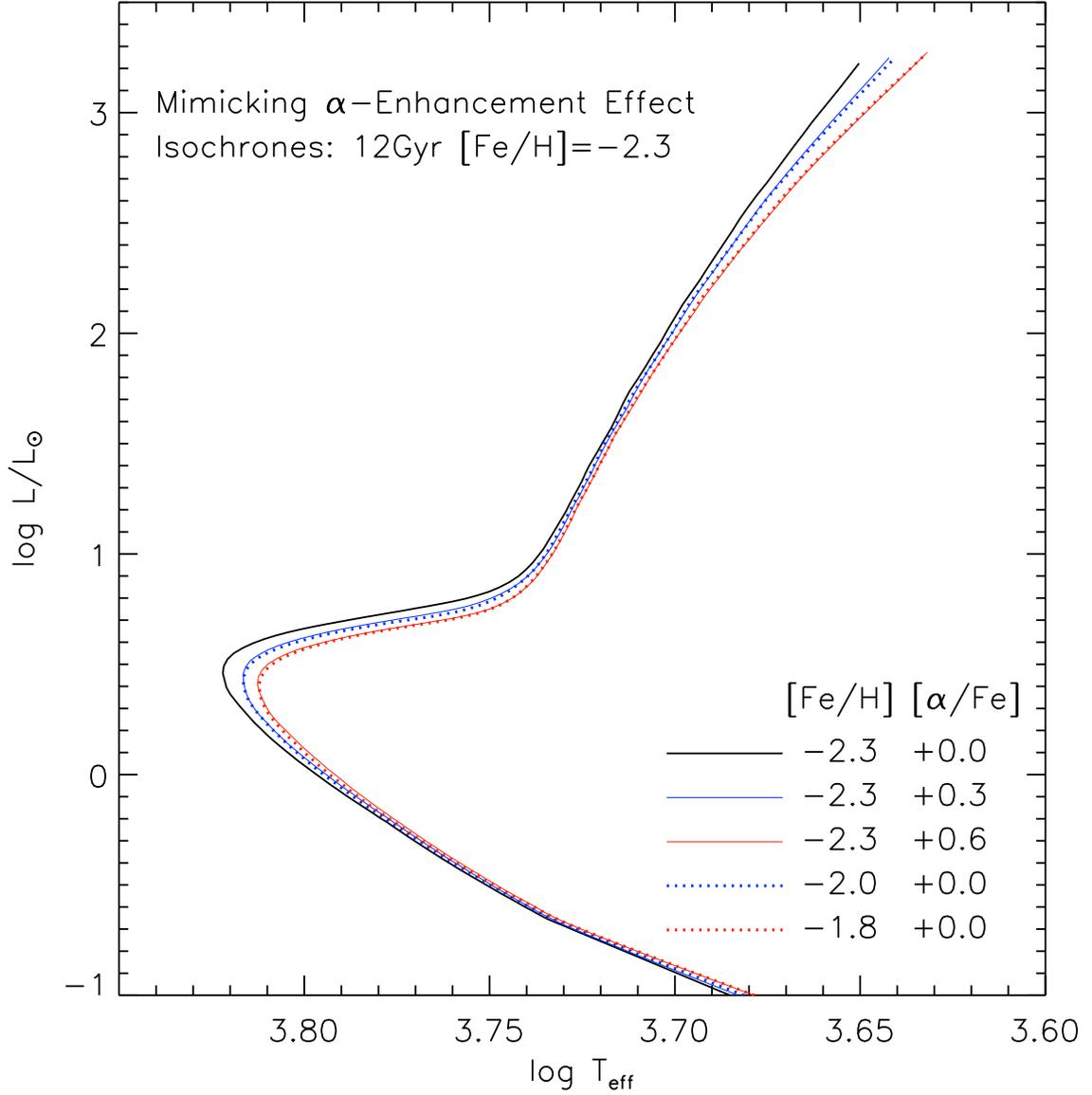}
\caption[f7.eps]{
The adequacy of the $\alpha$-enhancement effect simulation formulae
for the 12\,Gyr [Fe/H]=$-2.3$ isochrone. The left-most isochrone is 
non-$\alpha$-enhanced isochrone. The two continuous-line isochrones
on the right are $\alpha$-enhanced isochrones for the same [Fe/H].
They are well reproduced by more metal-rich, non-$\alpha$-enhanced isochrones
(dotted curves). The largest weight for the fit was given to the
main sequence.
\label{f7}}
\end{figure}

\begin{figure}
\plotone{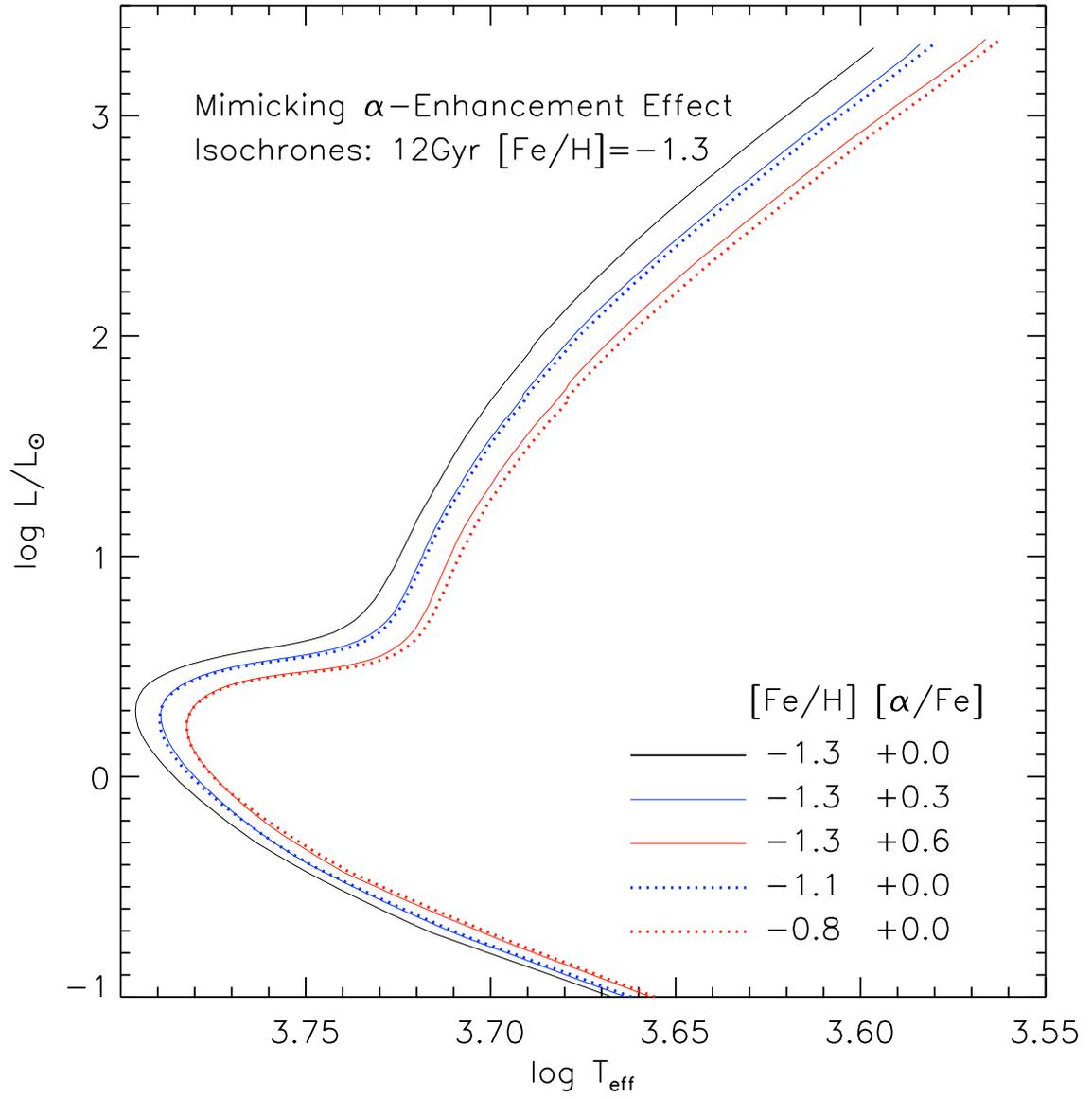}
\caption[f8.eps]{
Same as Figure \ref{f7} but for [Fe/H]=$-1.3$.
\label{f8}}
\end{figure}

	Figures \ref{f7} and \ref{f8} 
represent independent confirmation of the results of \citet{Van00}.
The plots
show that at least at low metallicities the $\alpha$-enhancement effect on the
overall shape of the isochrone can be mimicked by adopting more
metal-rich non-$\alpha$-enhanced isochrones.
	{\em It is important to note that our matching effort was made
to fit the upper MS}.
	Since empirical determination of the turn-off temperature 
is not easy, and since the turn-off luminosities of the simulated isochrones
seem not so different from those of the $\alpha$-enhanced ones (see below), 
it seemed to us sensible to concentrate on the upper MS near turn-off, but
not on the turn-off itself. 

	Figure \ref{f7} shows the 12\,Gyr isochrones 
for [Fe/H]=$-2.3$.
	The twice and four times $\alpha$-enhanced models are compared 
with the more metal-rich non-$\alpha$-enhanced models.
	The overall fit is good.
	It is clear that as metallicity increases, such scalings 
become less reliable.
	At [Fe/H]=$-1.3$, the simulated models (dotted lines)  
produce good fits on the MS of the $\alpha$-enhanced models but 
fail to match the giant branches, as shown in Figure \ref{f8}. 
	A bit of the shift in the giant branches may be corrected by a 
small reduction of $Z$.
	But in this analysis, we assumed that the MS is generally the 
branch that is best determined empirically and thus put more weight on
MS fitting.
	Besides, such small differences will be considered smaller than
the accuracy of current stellar models anyway.

\begin{figure}
\plotone{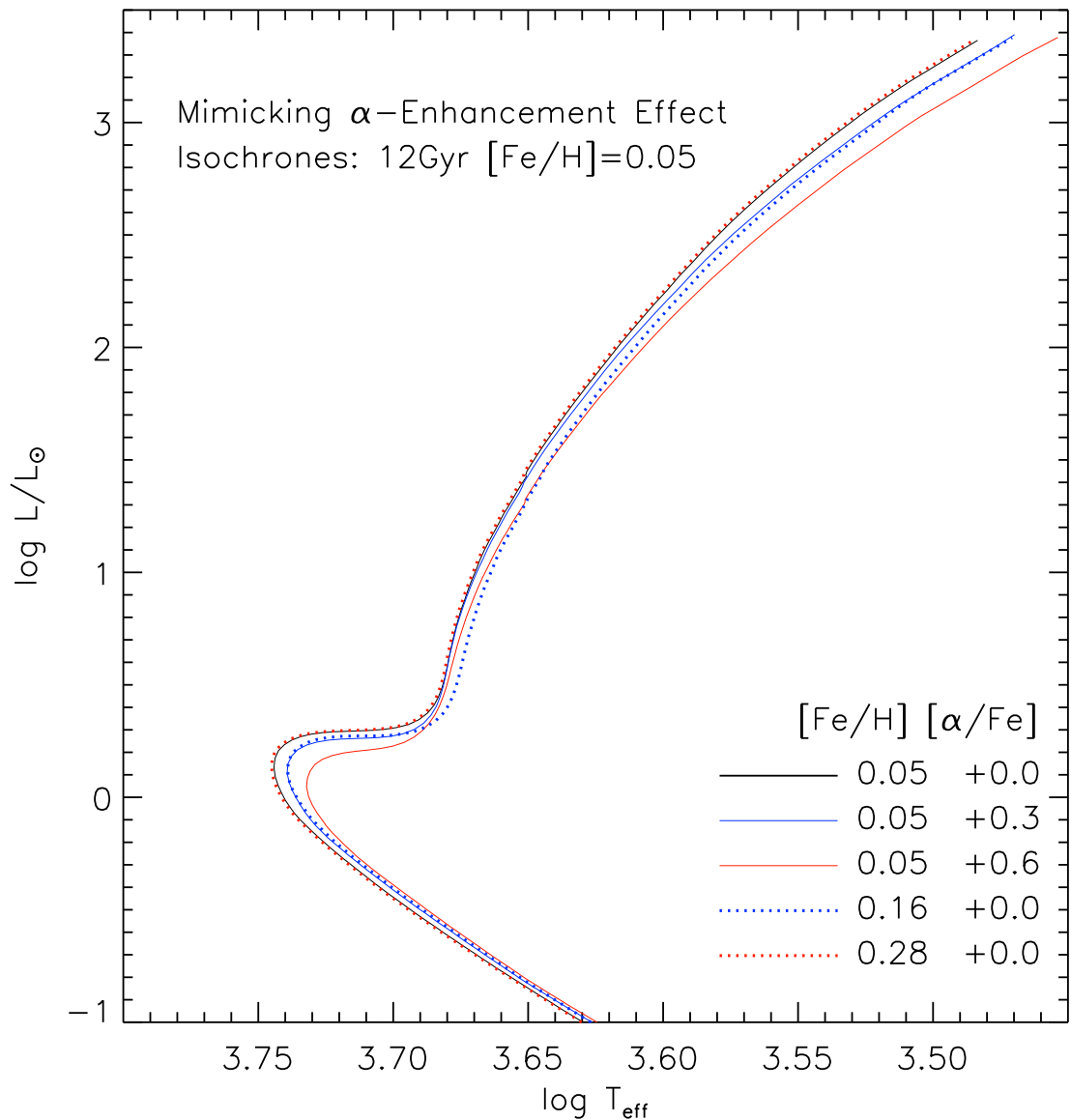}
\caption[f9.eps]{
Same as Figure \ref{f8} but for [Fe/H]=0.05.
At such high metallicities, the simulation formulae do not perform well
at all. Even if we find different formulae for high metallicities,
there is no way of reproducing good fits on the main sequence and the
giant branch simultaneously.
\label{f9}}
\end{figure}

	The fits become impossible when metallicity is as high as close
to solar, as shown in Figure \ref{f9}.
	Our numerous simulations showed that no single simulated
isochrone can match the true $\alpha$-enhanced model for all of MS,
MS turn-off, and red giant branch simultaneous.
	We conclude that it is $not$ recommended to use any
such simulation formulae at high metallicities.

\begin{figure}
\plotone{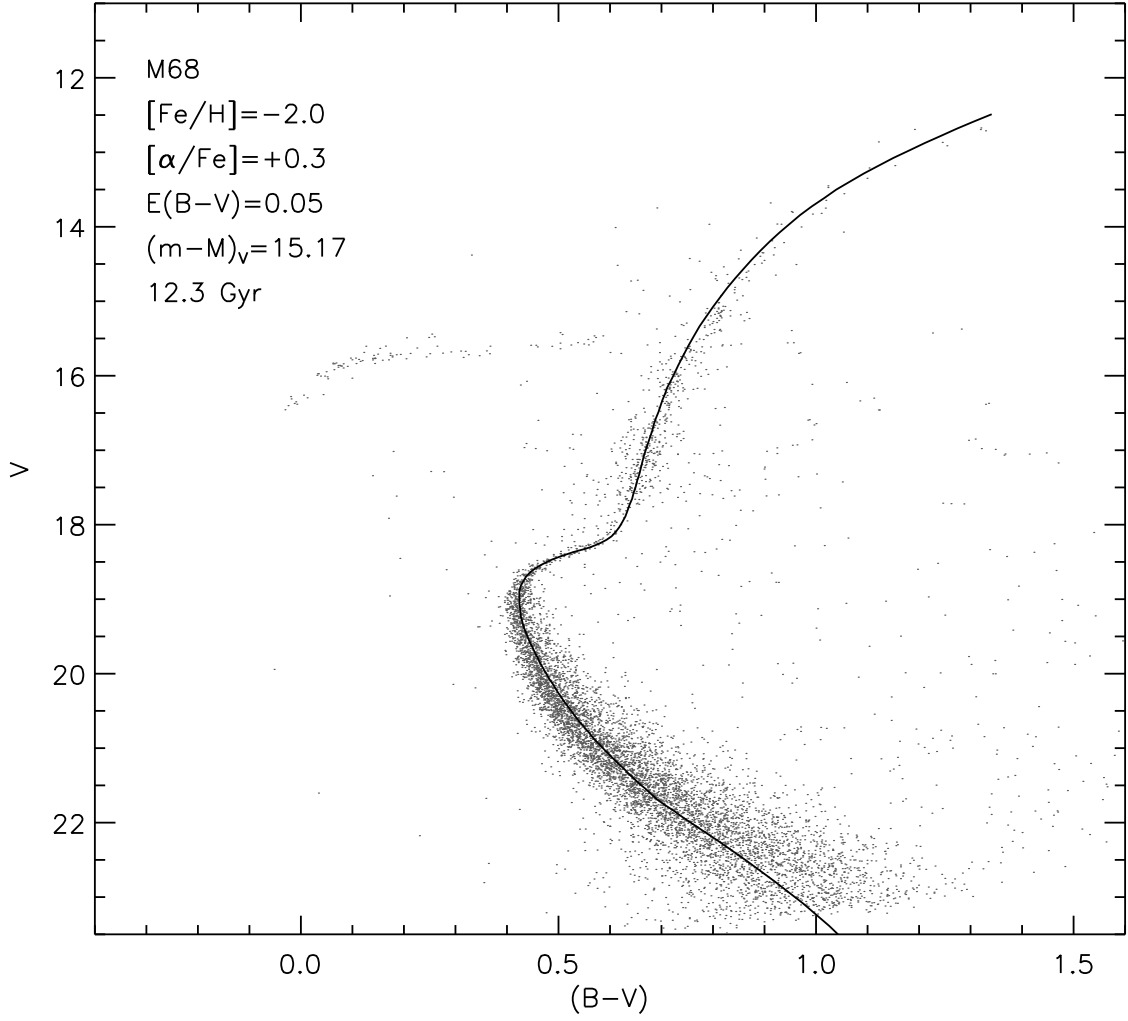}
\caption[f10.eps]{
A sample fit to the metal-poor cluster M\,68 using the $\alpha$-enhanced 
$Y^{2}$ Isochrone. The CMD data are from \citet{Wal94}.  The rough
estimates of reddening and metallicity are from \citet{Har96}.
\label{f10}}
\end{figure}

\begin{figure}
\plotone{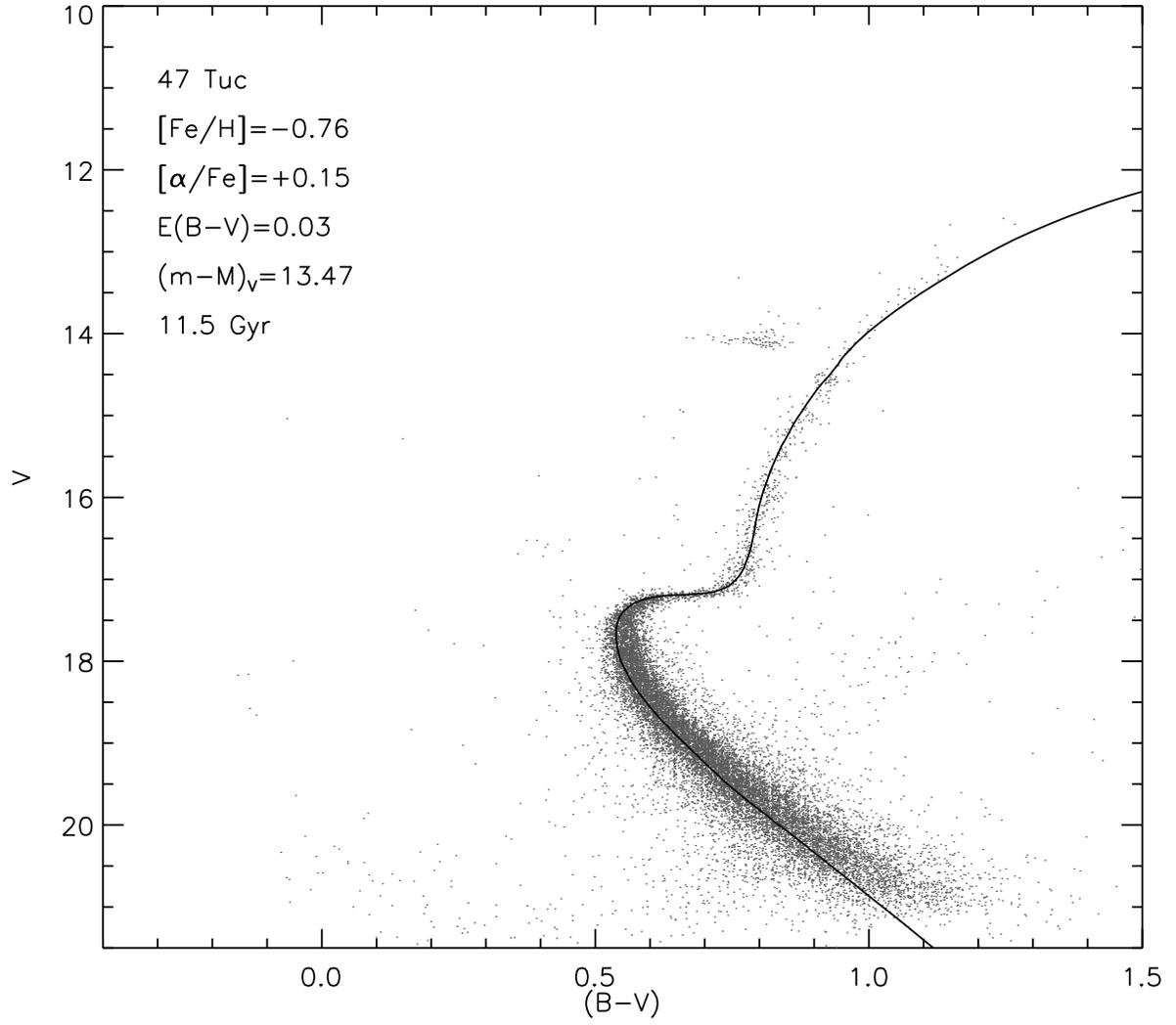}
\caption[f11.eps]{
A sample fit to the metal-rich cluster 47\,Tuc using the $\alpha$-enhanced 
$Y^{2}$ Isochrone. The CMD data have been kindly provided by A. Sarajedini.
\label{f11}}
\end{figure}

       In Paper 1, we presented isochrone fits to two globular
clusters, one metal-poor (M\,68), the other metal-rich (47\,Tuc),
 using non-$\alpha$-enhanced isochrones but corrected for
$\alpha$-enhancement using previously available simulation formulae.
       We present the same fits now using true $\alpha$-enhanced
isochrones.
       Figures \ref{f10} and \ref{f11} show the fits.
       The fits are at least as good as in Paper 1; the only
difference is that we do not need such formulae any more.
In general, globular cluster distances are still uncertain 
that one should not place too much reliance on the quality
of this kind of CMD fitting.
It is remarkable, however, that the distance modulus
used for 47\,Tuc in Figure \ref{f11}, (m-M)=13.47,
is within 1.0 $\sigma$ of the latest value based on
the main sequence fitting using 43 local subdwarfs
(Hipparcos) which is $m-M(V)=13.37^{+ 0.10}_{ -0.11}$ \citep{Per02}.
And, it is within 1.4 $\sigma$ of
the white dwarf distance modulus, 13.27 $\pm$ 0.14
\citep{Zoc01}.
This is a rather good agreement
considering many uncertainties not only in distance, but also in the
bolometric correction and color calibration etc.

\section{Summary and discussion}

	This paper presents an extension of the $Y^{2}$ isochrones 
\citep{paper1} including the effects of $\alpha$-element enhancement. 
	Two values of $\alpha$-enhancement, [$\alpha$/Fe]=$+$0.3 and 
$+$0.6 have been considered.

	As in Paper 1, pre-MS 
evolutionary phases are included, allowing for the 
construction of younger isochrones.  
	Isochrones including the complete hydrogen burning phase, 
suitable for population synthesis, have been constructed for the age 
range 0.1 -- 20\,Gyr.  
	In the case of stars with a convective core near the MS, 
overshoot by 0.2 pressure scale height at the edge of the convective 
core has been taken into account.
	An additional set of very young isochrones for ages 1 -- 80\,Myr is 
presented.
	They are unsuitable for population synthesis because their 
post-MS phases are not included but would be useful for comparing with the 
CMDs of very young stellar populations. 

	We confirm that the use of simple scaling formulae, such as the one
introduced by Salaris et al.~(1993), can approximate the effect of 
$\alpha$-enhancement on the isochrone morphology especially at low 
metallicities.
	Thus, such formulae, in the absence of appropriate models,
can be useful to the studies of globular clusters 
and dwarf spheroidal galaxies.    
	However, the range of validity of such scaling laws is limited.  
	In general, because of the complex interaction between opacities, 
the equation of state and nuclear processes, it is not possible to derive 
a scaling formula that is applicable for a large range of $Z$.  

	From the point of view of stellar population chronology, 
this paper confirms that for a given [Fe/H],
the updated isochrones that include the effects of
 $\alpha$-element enhancement 
lead to significant age reductions of up to 23\% if [$\alpha$/Fe]=$+$0.3, 
as shown in Table \ref{age}.
Approximately a third of this reduction is due to a factor 
of two increase in the $\alpha$-element abundances
(see the difference between the second and fifth columns in Table \ref{age}).
	A new result, which is equally important, is the age increase at high 
$Z$, which removes the age gap that used to exist between the oldest 
metal-rich and metal-poor stellar populations.  

	Caveats of this work include the treatment of $\alpha$-enhancement
in the temperature-color transformation.
	Theoretical stellar properties (temperature, gravity, etc.)
are converted to observable quantities (magnitudes and colors) based on
a stellar spectral library (in our case, either the Green et al. table or
the Lejeune et al. table).
	Such a spectral library is usually calibrated empirically so that
it reproduces the colors of sample stars in the Milky Way 
for a given set of parameters.
In fact, the calibration is based on a limited and incomplete sample.
	For example, such Milky Way samples may imply 
[$\alpha$/Fe]=0.3 for [Fe/H]=$-$2, [$\alpha$/Fe]=0.15 for [Fe/H]=$-$0.7, and
[$\alpha$/Fe]=0.0 for [Fe/H]=0.
Users of these isochrones should be warned that for this reason, interpolated 
isochrones, even if well within
our parameter range ([$\alpha$/Fe] = 0.0 -- 0.6), are not necessarily
based on a 
proper temperature-color transformation scheme.
	Some colors are more sensitive to [$\alpha$/Fe] than others.
	Thus, a full $\alpha$-enhancement treatment will be possible only when
temperature-color transformation schemes are available for a wide
range of $\alpha$-enhancement as well.

	Similarly, the importance of $\alpha$-enhancement effects on 
the spectral line analysis for clusters and galaxies has been recognized 
\citep{Vaz01} as well.
	The effects of $\alpha$-enhancement occur mainly in two steps.
	The first is in the stellar model construction through opacities 
and energy generations.
	The next is through the convolution of the stellar models with
a stellar spectral library.
	For this reason, one must have a properly $\alpha$-enhanced
stellar spectral library for a wide range of $\alpha$-enhancement 
in order to generate consistent population synthesis models.

	Despite the caveats, we believe that these new isochrones with
$\alpha$-enhancement options will be useful in many studies.
	The full set of isochrones and a FORTRAN package that work
for age, metallicity, and $\alpha$-enhancement interpolation are
available at the following internet sites.

{\tt http://www-astro.physics.ox.ac.uk/$\sim$yi/yyiso.html}

{\tt http://www.astro.yale.edu/demarque/yyiso.html}

{\tt http://csaweb.yonsei.ac.kr/$\sim$kim/yyiso.html}

\acknowledgments
We are indebted to A. Sarajedini for providing the CMD data for 47\,Tuc.
This research has been supported by Korean Research Foundation Grant
KRF-2001-003-D00106 (YCK), and received partial support from 
NASA grant NAG5-8406 (PD).


\clearpage

\end{document}